\documentclass[12pt]{article}
\title{Box model for channels of human migration}
\author{Nikolay K. Vitanov$^{1,2}$, Kaloyan N. Vitanov$^1$}
\date{$^1$Institute of Mechanics, Bulgarian Academy of Sciences, Acad. G. Bonchev Str., Bl. 4, 1113 Sofia, Bulgaria \\
$^2$Max-Planck Institute for the Physics of Complex Systems, N{\"o}thnitzerstr. 36,
01187, Dresden, Germany}
\begin{document}
\maketitle
\begin{abstract}
We discuss a mathematical model of migration channel based on the truncated Waring distribution. The truncated Waring distribution is obtained for 
a more general model of motion of substance through a channel containing 
finite number of boxes. The model is applied then for case of migrants moving through a channel consisting of finite number of countries or cities. The number of migrants in the channel strongly depends on the number of migrants that enter the channel through the country of entrance. It is shown that if the final destination country is very popular then large percentage of migrants may concentrate there.
\end{abstract}
\section{Introduction}
Population migration involves the relocation of individuals, households or moving groups between geographical locations. Much efforts are directed to the study of internal migration in order to understand this migration and to
make projection of the internal migration flows that may be very important for
taking decisions about economic development of regions of a country (Armitage (1986), Braken and Bates (1983), Champion et al. (2002)). In addition the study 
of migration becomes very actual after the large migration flows directed to Europe in September 2015. From the point of view of migrating units
migration models may be classified as 
macromodels or micromodels (Stillwell and Congdon (1991), Cadwallader (1989, 1992)).
Micromodels are based on the individual migrating unit (person, group
or household) and on the processes underlying the decision of the migrant to remain in the current location or to move somewhere else (Maier and Weiss (1991)).
Macromodels are for the aggregate migration flows. Many macromodels are explanatory and they use non-demographic information. An important class of these models are the gravity models  having their roots in the ideas of Ravenstein about the importance of distance for migration and their first applications in the study of intercity migration by Zipf (Ravenstein (1885), Zipf (1946)). Further development of these models was made by  the concept of spatial interactions (Wilson (1970), Stillwell (1978), Fotheringham et al. (2001)) or by introduction of statistical spatial interaction models (Congdon (1991), Flowerdew (1991), 
Nakaya (2001), Fotheringham et al. (2002)).
\par
Other kinds of macromodels use demographic information from the field of
multi-state demography in order to generate projections of migration flows. In the first models of this kind one used a cohort component model which
involved the estimation of the population in the studied region  at the
beginning of a projection period. Then a projection of the number of births during the future time period and the survival of those in existence or being born during the period was made (Bowley (1924), Whelpton(1936)). In the course of the  time these models become multi-regional stochastic models and  the requirement to the used information was changed: the first  models required little information about migration and the more sophisticated models  require maximum information about migration, e.g. a migration flow information disaggregated by single year of age and sex (Rees and Wilson (1977), Rogers (1990, 1995)).
\par 
Migration models may be classified as probability models 
or deterministic models with respect to their mathematical features. Many probability models of migration 
(Willekens  (1999, 2008)) are 
focused on the change of address in the process of migration involving the 
crossing of an administrative boundary. The migration data can be connected to
the events of migration (event data or movement data) or to the place of residence (migrant status) at a given point of time (status data). Interesting
kind of data are the duration data where the time is measured as a duration of since a reference event (event-origin that may be associated with the
start of the process of  migration). The observation of place of residence of an individual at two points in time leads to collection of transition data (Ledent (1990), Willekens (1999)). Data
obtained by recording the migrant status at several points of time are called panel data. 
\par
Different kinds of data are connected to different probability models.
The \emph{duration data} may be used in the \emph{exponential model of 
migration} (Willekens (1999)). This model considers two states - origin state 
and a destination state. It is assumed that at the onset of migration all individuals are in the origin state. Individuals may leave the origin state for different reason but only the leaving connected to migration are considered. 
And the event of migration is assumed to be experienced by an individual only once (non-repeatable event). Let the size of population (of
identical individuals) be $m$ and let $T$ be the time at which an individual migrates (T=0 denotes the birth of the individual). One can define probability
distribution $F(t)=P(T<t)$ ($T<t$ means that the migration event happens between
$T=0$ and $T=t$). $f(t)$ is the probability density function connected to 
$F(t)$ and $S(t)=1-F(t)$ is called survival function. It is connected to the 
size $mS(t)$ of the risk set of individuals (these who have not migrated but are
exposed to the risk of migration in the future). One can introduce the conditional density function $\mu(t)=f(t)/S(t)$ which represents probability that the migration event occurs in a small interval following $t$ provided that it was not occurred before $t$. The survival function and the probability density functions then are
\begin{equation}\label{models1}
S(t) = \exp \left[ - \int \limits_{0}^t dt \mu(t) \right]; \ \ f(t) =\mu(t) \exp \left[ - \int \limits_{0}^t dt \mu(t) \right].
\end{equation}  
If $\mu(t)$ is constant (equal to $\mu$) it is referred to as a migration rate.
Migration rate $\mu$ can be determined on the basis of an
appropriate likelihood function (Willekens (1999)). The exponential model of
migration can be extended in different directions, e.g, to the case of non-identical individuals; the life span can be split into age intervals (of one or several years) (Blossfeld and Rohwer (2002), Blossfeld et al. (2007)); the number of possible destination states may become larger than 1 (Hachen (1988)).
\par
The exponential model of migration is obtained on the basis of assumption that 
migration is a non-repeatable event. Let us now assume that the migration is a 
repeatable event and there is no upper limit on the number of migrations in
a time interval. If the event data are available the \emph{Poisson model} of migration can be used. The Poisson  model  describes the number of migrations during an interval of unit length(e.g. year, month, etc.). In this model the probability of 
observing $n$  migrations during the unit interval is given by the Poisson distribution
\begin{equation}\label{add1}
P(N=n) = \frac{\lambda^n}{n!} \exp(-\lambda)
\end{equation}
where $\lambda$ is the expected number of migrations during the unit interval.
\par
As we have mentioned above the Poisson model is applicable when event data are available. When status data for migration are available, i.e., migration is measured by comparing 
the places of residence at two consecutive points of time, then the
probability of observing $n$ migrants among a sample population of $m$
individuals is given by the binomial distribution (\emph{binomial model of 
migration})
\begin{equation}\label{add2}
P(N=n) = \frac{m!}{n!(m-n)!}p^n(1-p)^{m-n}
\end{equation} 
of index $m$ and parameter $p$ representing the probability of being migrant.
If the number of possible destinations for migration is larger than $1$ (let us
assume that the number of possible destinations be $K$) then the probability 
distribution connected to migration is given by the multinomial distribution
(\emph{multinomial model of migration})
\begin{equation}\label{add3}
P(N_1=n_1,\dots,N_K=n_K) = \frac{m! \prod \limits_{i=1}^K p_i^{n_i}}{\prod 
\limits_{i=1}^K n_i!}; \ \sum \limits_{i=1}^K p_i =1; \sum \limits_{i=1}^K n_i 
= m
\end{equation}
The transition between the states (e.g. addresses occupied in different ages 
is given by transitional probabilities $p_{ij}(x)$ which correspond to the
probability that an individual who resides in state $i$ at $x$ resides in 
state $j$ at $t+1$. $p_{ij}$ may be represented as (Willekens (2008))
\begin{equation}\label{add4}
p_{ij}(x) = \frac{\exp[\beta_{j0}(x) + \beta_{j1}(x) Y_i(x)]}{
\sum \limits_{r=1}^K \exp[\beta_{j0}(x) + \beta_{j1}(x) Y_r(x)]}
\end{equation}
for the case when only the most recent state occupancy is relevant and the
person has single relevant attribute (covariate) $Y_i$ that is equal to $1$
if the state is occupied and is equal to $0$ otherwise.
Then the state probability $\pi_i$ that an individual occupies state $i$
is given by 
\begin{equation}\label{add5}
\textrm{logit} (\pi_j) = \ln \left(\frac{\pi_j}{1-\pi_j} \right) =
 \beta_{j0}(x) + \beta_{j1}(x) Y_i(x)
\end{equation}
From the transition probabilities for discrete time processes one may turn
to transition rates and this way leads to the Markov chain models (Singer and Spilerman (1979)). The parameters of these models may be estimated even if some data are missing (McLachlan and Krishnan (1997)). Markov chain models are 
useful  to demographers concerned with problems of movement of people 
(Collins (1972, 1975)) as these models are  appropriate for describing and
analysing the nature of changes generated by the movement from one state 
of a system to another possible state and in some cases Markov models may be 
useful for  forecasting future changes. Bayesian modelling framework may be used for generating estimates of place-to-place migration flows too (Raymer (2007), Bierley et al. (2008)). 
\par
From the deterministic models  we shall give brief additional information
about the gravity models and about models connected to the replicator dynamics. 
Finally we shall mention some urn models of interest for our study.
\par
The gravity model of migration (already mentioned above in the text) is a place-to-place migration model that assumes that interregional migration is directly related to the population of the
origin and of the destination regions and inversely related to the distance between them (Greenwood (2005)). The classic gravity model can be written as
\begin{equation}\label{grav1}
\ln(M_{ij}) = \ln(G) + \beta_1 \ln(P_i) + \beta_2 \ln(P_j) - \alpha \ln(D_{ij})
\end{equation}
where $M_{ij}$ is the migration from region $i$ to region $j$; $G$ is a 
constant; $P_{i}$ and $P_{j}$ are populations of regions $i$ and $j$ and 
$D_{ij}$ is the distance between the regions. The gravity model may be 
extended in different ways, e.g., to include the income and unemployment in 
the two regions as well as in the ways discussed above in the text.
\par
A simple version of the replicator equation describes the time change 
of the part $p_i$ of the
population of $n$ individuals ($ p_i = n_i/n$) that use strategy $s_i$ at
time $t$ ($n = \sum \limits_{i=1}^m n_i$) for the case of symmetric game
(Cressman and Tao (2014), Karev and Kareva (2014)). The form of the replicator equation is 
\begin{equation}\label{repl1}
\frac{dp_i}{dt} = p_i[\pi(s_i,p) - \pi(p,p)], \ i=1,\dots,m
\end{equation}
where $\pi(s_i,p)$ $\pi(p,p) = \sum \limits_{j=1}^m p_i \pi(s_i,p)$ are
the payoff of the individuals that follows the strategy $s_i$ and the
the population mean payoff.
\par
The urn models that may be studied analytically usually consider a single urn 
containing balls of different colours and a fixed set of rules that specify
the way the urn composition evolves: at each discrete instant, a ball is 
picked up at random, its color is inspected, and, in compliance with the 
rules, a collection of balls of various colours is added to the urn (Flajolet
et al.(2006)). The balls may be exchanged between two or more urns - an example of such model is the Bernoulli-Laplace model for exchange of balls between two urns or the model of transfer of balls from one urn to another one 
(Blom et al. (1994)). As we shall see below our model possesses some
features of replicator model as well as of an urn model. 
\par
Our interest in
the problems of human migration arose in the course of the research on 
ideological struggles (Vitanov et al. (2010, 2012)) and waves and statistical 
distributions in population and other  systems (Vitanov et al. (2009a,b,2011, 2013a, 2013b, 2015), Vitanov and 
Vitanov (2013)), Vitanov and Dimitrova (2010, 2014). Migration channels are one possibility for movement of 
migrants. Below we shall discuss a mathematical model of migration channels
based on the  truncated Waring distribution.
The short review of the models above allows us to mention the place of our 
model within the set of models of migration. 
Our model is a multi-regional macromodel. In addition we
shall observe that the discussed model doesn't belong to the class of
the probability models of migration despite the fact that it is closely
connected to the Waring distribution. We shall see that: (i) From the point 
of view of the model equations our model is a deterministic model
that is close to the class of replication models and to the sub-class of the
replication-mutation models; (ii) From the point of view of distribution of
migrants in the countries of the channel our model may be considered
as an urn model.
\par 
The text below is organized as follows. In Sect.2 the truncated Waring distribution is obtained as a result of mathematical model of  a substance that moves through a finite sequence of cells. In Sect. 3 the discussed model is applied to the channels of migration where the
moving substance are the migrants and the sequence of cells are countries or
cities that are on the route of migration. A discussion on: (i) the place of
our models in the set of models of migration; (ii) the meaning of the
parameters of the model, and (iii) application of the model for purposes other
than explanatory ones,  is presented in Sect.4.
\section{The model and the truncated Waring distribution}
Let us consider the following simple model (Schubert and Gl{\"a}nzel (1984)).
We have an  array of $N+1$ cells (boxes) indexed in succession by
non-negative integers, i.e., the first cell has index $0$ and the last
cell has index $N$. In the model discussed by Schubert and Gl{\"a}nzel
$N$ is infinite. We shall treat $N$ as finite number and this will
lead us to an important effect (concentration of migrants in the final destination country). We assume that
there exists an amount $x$ of some substance that is
distributed among the cells. Let $x_i$ be the amount of the substance in the
$i$-th cell. Then
\begin{equation}\label{warig1}
x = \sum \limits_{i=0}^N x_ i
\end{equation}
The fractions $y_i = x_i/x$ can be considered as probability values of
distribution of a discrete random variable $\zeta$
\begin{equation}\label{warig2}
y_i = p(\zeta = i), \ i=0,1, \dots, N
\end{equation}
The expected value of the random variable $\zeta$ is 
\begin{equation}\label{warig3}
E(\zeta) = \sum \limits_{i=0}^N i y_i
\end{equation}
\par
The amount $x_i$ can change due to the following 3 processes:
\begin{enumerate}
\item Some amount $s$ of the substance $x$  may enter the system of
cells from the external environment through the $0$-th cell;
\item Amount $f_i$  may be transferred from the $i$-th
cell into the $i+1$-th cell;
\item Amount $g_i$  may leak out the $i$-th cell into the
external environment.
\end{enumerate}
The above processes can be modeled mathematically by the system of ordinary
differential equations:
\begin{eqnarray} \label{warig4}
\frac{dx_0}{dt} &=& s-f_0-g_0; \nonumber \\
\frac{dx_i}{dt} &=& f_{i-1} -f_i - g_i, \ i=1,2,\dots, N-1 \nonumber \\
\frac{dx_N}{dt} &=& f_{N-1}  - g_N .
\end{eqnarray}
The following forms of the amounts of the moving substance are assumed in 
(Schubert and Gl{\"a}nzel (1984)) ($\alpha, \beta, \gamma, \sigma$ are 
parameters)
\begin{eqnarray}\label{warig5}
s &=& \sigma x; \ \ \sigma > 0 \to \textrm{self-reproducing property} \nonumber \\
f_i &=& (\alpha + \beta i) x_i; \ \ \ \alpha >0, \ \beta \ge 0 \to
\textrm{cumulative advantage of higher cells} \nonumber \\
g_i &=& \gamma x_i; \ \ \ \gamma \ge 0 \to \textrm{uniform leakage over the cells}
\end{eqnarray}
Substitution of Eqs.(\ref{warig5}) in Eqs.(\ref{warig4}) leads to the
relationships
\begin{eqnarray}\label{warig6}
\frac{dx_0}{dt} &=& \sigma x - \alpha x_0 - \gamma x_0; \nonumber \\
\frac{dx_i}{dt} &=& [\alpha+ \beta(i-1)]x_{i-1} - (\alpha + \beta i +\gamma)x_i,
\ i=1,2,\dots,N-1
\nonumber \\
\frac{dx_N}{dt} &=& [\alpha+ \beta(N-1)]x_{N-1} - \gamma x_N
\end{eqnarray}
\par
Let us sum the equations from (\ref{warig6}). The result of the summation is
\begin{equation}\label{warig6a}
\frac{dx}{dt} = (\sigma - \gamma) x
\end{equation}
and the solution for $x$ is
\begin{equation}\label{warig7}
x = x(0) \exp[(\sigma - \gamma)t]
\end{equation}
where $x(0)$ is the amount of $x$ at $t=0$.
\par
The distribution of $y_i$ will lead us to the truncated Waring
distribution. From Eqs.(\ref{warig6}) and with the help of Eq.(\ref{warig7})
and the relationship $\frac{dy_i}{dt} = \frac{1}{x^2}\left[ x \frac{dx_i}{dt}  -
x_i \frac{dx}{dt}\right]$ one obtains
\begin{eqnarray}\label{warig8}
\frac{dy_0}{dt}&=& \sigma -(\alpha + \sigma) y_0; \nonumber \\
\frac{dy_i}{dt}&=&[\alpha+\beta(i-1)]y_{i-1} - (\alpha + \beta i + \sigma) y_i,
\ i=1,2,\dots,N-1
\nonumber \\
\frac{dy_N}{dt}&=&[\alpha+\beta(N-1)]y_{N-1} -  \sigma y_N
\end{eqnarray}
We search for  solution of Eq.(\ref{warig8}) in the form
\begin{equation}\label{warig9}
y_i = y_i^* + F_i(t)
\end{equation}
where $y_i^*$ is the stationary solution of Eqs.(\ref{warig9}) given
by the relationships
\begin{eqnarray}\label{warig10}
y_0^* &=& \frac{\sigma}{\sigma + \alpha} \nonumber \\
y_i^* &=& \frac{\alpha + \beta(i-1)}{\alpha + \beta i + \sigma} y_{i-1}^*, \ i=1,2,\dots, N-1; \nonumber \\
y_N^* &=& \frac{\alpha+\beta(N-1)}{\sigma} y_{N-1}^*
\end{eqnarray}
For the functions $F_i$ we obtain the system of equations
\begin{eqnarray}\label{warig11}
\frac{dF_0}{dt}&=& -(\alpha + \sigma) F_0; \nonumber \\
\frac{dF_i}{dt}&=&[\alpha+\beta(i-1)]F_{i-1} - (\alpha + \beta i + \sigma) F_i,
\ i=1,2,\dots,N-1
\nonumber \\
\frac{dF_N}{dt}&=&[\alpha+\beta(N-1)]F_{N-1} -  \sigma F_N
\end{eqnarray}
The solutions of these equations are
\begin{equation}\label{sol1}
F_0(t) = b_{00} \exp[-(\alpha + \sigma)t]
\end{equation}
\begin{equation}\label{sol2}
F_1(t) = b_{10} \exp[-(\alpha + \sigma)t] + b_{11} \exp[-(\alpha + \beta + \sigma)t]
\end{equation}
\begin{center}
$\dots$
\end{center}
\begin{equation}\label{soli}
F_i(t) = \sum \limits_{j=0}^i b_{ij} \exp[-(\alpha + \beta j + \sigma)t]; 
\ i=1,2,\dots,N-1
\end{equation}
\begin{equation}\label{soln}
F_N(t) = \sum \limits_{j=0}^N b_{Nj} \exp[-(\alpha + \beta j + \sigma)t].
\end{equation}
where
\begin{eqnarray}\label{soly}
b_{ij} &=& \frac{\alpha + \beta(i-1)}{\beta(i-j)} b_{i-1,j}; \ i=1,\dots,N-1;
\ j=0,\dots, i-1;
\nonumber \\
b_{Nj} &=& - \frac{\alpha + \beta(N-1)}{\alpha + j \beta} b_{N-1,j}, \ j=0, \dots, N-1
\nonumber \\
b_{NN} &=& 0.
\end{eqnarray}
$b_{ij}$ that are not determined by Eqs.(\ref{soly}) may be determined by the initial conditions. In the exponential
function in $F_i(t)$ there are no negative coefficients and because of this 
when $t \to \infty$  $F_i(t) \to 0$ and the systems comes to the
stationary distribution from Eqs.(\ref{warig10}). The form of this
stationary distribution is:
\begin{eqnarray}\label{warig12}
P(\zeta = i) &=& \frac{a}{a+k} \frac{(k-1)^{[i]}}{(a+k)^{[i]}}; \ \ k^{[i]} = \frac{(k+i)!}{k!}; \ i=0,\dots,N-1 \nonumber \\
P(\zeta = N) &=& \frac{1}{a+k} \frac{(k-1)^{[N]}}{(a+k)^{[N-1]}},
\end{eqnarray}
with parameters $k = \alpha/\beta$ and $a=\sigma/\beta$.
\par
The obtained distribution is close to the Waring distribution that can be obtained for the case of channel with infinite number of cells (see Appendix 1).
The distribution (\ref{warig12}) has a concentration of substance in the last cell (i.e. in the $N$-th cell). For the case of Waring distribution the same
substance is distributed in the cells $N$, $N+1$, $\dots$. 
\par
Let us calculate one example connected to the obtained distribution.
Let us have 6 cells ($N=5$). Then
\begin{eqnarray}\label{waring13}
P(0) &=& \frac{\sigma}{\sigma + \alpha}; \nonumber \\
P(1) &=&  \frac{\alpha}{\alpha + \beta + \sigma} \frac{\sigma}{\sigma + \alpha};
\nonumber \\
P(2) &=& \frac{\alpha + \beta}{\alpha + 2 \beta + \sigma} \frac{\alpha}{\alpha + \beta + \sigma} \frac{\sigma}{\sigma + \alpha};\nonumber \\
P(3) &=& \frac{\alpha + 2 \beta}{\alpha + 3 \beta + \sigma}\frac{\alpha + \beta}{\alpha + 2 \beta + \sigma} \frac{\alpha}{\alpha + \beta + \sigma} \frac{\sigma}{\sigma + \alpha};\nonumber \\
P(4) &=& \frac{\alpha + 3 \beta }{\alpha + 4 \beta + \sigma}\frac{\alpha + 2 \beta}{\alpha + 3 \beta + \sigma}\frac{\alpha + \beta}{\alpha + 2 \beta + \sigma} \frac{\alpha}{\alpha + \beta + \sigma} \frac{\sigma}{\sigma + \alpha};\nonumber \\
P(5) &=& \frac{\alpha + 4 \beta}{\sigma + \alpha} \frac{\alpha + 3 \beta }{\alpha + 4 \beta + \sigma}\frac{\alpha + 2 \beta}{\alpha + 3 \beta + \sigma}\frac{\alpha + \beta}{\alpha + 2 \beta + \sigma}  \frac{\alpha}{\alpha + \beta + \sigma} \nonumber \\
\end{eqnarray}
Let now we assume that $\alpha = \sigma/2$ and $\beta = \sigma/4$. Then
\begin{eqnarray}\label{waring14}
P(0) = \frac{2}{3}; \ P(1) = \frac{4}{21}; \ P(2) = \frac{1}{14}; \
P(3) = \frac{2}{63}; \ P(4) = \frac{1}{63}; \ P(5) = \frac{1}{42}. \nonumber \\ 
\end{eqnarray}
We note that $P(5) > P(4)$, i.e there can be a concentration of substance in the
last cell. This is difference with respect to the Waring distribution discussed
in Appendix A. We note also that the sum of all probabilities is equal to $1$
(as it can be expected).
\section{Application of the model to migration channels}
Let us consider a sequence of $N+1$ countries (or cities). A flow of migrants flows through this migration channel from the country of entrance to the final destination country. We can consider this sequence of countries as sequence of boxes (cells). The entry country will be the box with label $0$
and the final destination country will be the box with label $N$. 
Let us have  a number $x$ of migrants that are
distributed among the countries. Let $x_i$ be the number of migrants in the
$i$-th country. This number can change on the basis of the following three processes: (i) A number $s$ of migrants enter the channel  from the external
environment through the $0$-th cell (country of entrance); (ii)
A number $f_i$ of migrants can be transferred from the $i$-th country (city) to the $i+1$-th country (city); (iii)
A number $g_i$ from migrants change their status (e.g. they are not anymore
migrants and become citizens of the corresponding country).
\par
Let us assume that the number of migrants is large and continuum approximation may be used. Then the values of $x_i$ can be determined by Eqs. (\ref{warig4}).
The relationships (\ref{warig5}) mean that: (i) The number of migrants $s$ that 
enter the channel is proportional of the current number of migrants in all
countries (cities) that form the channel;  (ii)
There may be preference for some countries (cities), e.g. migrants may prefer
the countries that are around the end of the migration channel (and the
final destination country may be the most preferred one); (ii)
It is assumed that the conditions along the channel are the same with respect
to 'leakage' of migrants, e.g. the same proportion $\gamma$ of migrants 
leave the flow of migrants (e.g. they may become citizens of corresponding country, etc.)
\par
As it can be seen from Eq.(\ref{warig7}) the change of the number of migrants 
depends on the values of $\sigma$ (characteristic parameter for the migrants 
that enter the channel) and $\gamma$ (characteristic parameter for migrants 
that change their status. If $\sigma > \gamma$ the number of migrants in the 
channel increases exponentially. If $\sigma < \gamma$ the number of the 
migrants in the channel decreases exponentially. Thus there are three
main regimes of functioning of the channel.  
\par
The dynamics of the distribution of the migrants in the channel is modelled
by Eqs.(\ref{warig8}). When the time since the beginning of the operation
of the channel become large enough then the distribution of the migrants
in the cells of the channel (i.e. in the countries or cities that form the
migration channel) becomes close to the stationary distribution described
by Eqs. (\ref{warig10}). Let us stress that the stationary distribution described by (\ref{warig10}) is very similar to the Waring distribution but there is a substantial difference between the two distributions due to
the finite length of the migration channel: there may be large concentration
of migrants in the last cell of the channel (in the final destination country).
In order to illustrate this let us consider the case of large $\beta$ (final
destination country is very popular among the migrants). Let $\beta >> \alpha + \sigma$ and $\alpha >> \sigma$. Then $y_0^* \approx \sigma/\alpha \to 0$; $y_i^* \approx \sigma/(\alpha + \beta i) \to 0$ for $i=1,2,\dots,N-1$ and $y_N^* \approx 1$, i.e. almost all migrants may reach the final destination
country. Let us  refine this example by substituting numbers in Eqs.(\ref{waring13}). Let  $\alpha = 1/10$, $\beta=1$, $\sigma = 1/100$. Then
approximately in the entry country ($0$-th cell of the channel) there will be 
$9.09 \%$ of the migrants; in the following country (the first cell) there 
will be $0.82 \%$ of the migrants. In the third country (second cell) there 
will be $0.43 \%$ of the migrants. In the following country (third cell of 
the channel) there will be $0.29 \%$ of the migrants. In the  fifth country 
(fourth cell of the channel) there will be $0.22 \%$ of the migrants. In the 
final destination country there will be $89.15 \%$ of the migrants. Hence if 
the final destination country becomes very popular this  may lead to a large 
concentration of migrants there.
\section{Discussion}
The model of migration channels presented in sections 2 and 3  is connected 
to the Waring distribution. Nevertheless the model is not a probability model.
It is a deterministic model that leads to several results and one of these
results is a probability distribution. The deterministic results 
of the model are connected to the system of deterministic ordinary
differential equations. The probability result of the model is connected to the
fact that a migration channel is discussed. Thus more than 1 country
is considered and because of this a probability distribution connected to the
numbers of migrants in the countries of the channel  can be derived.
The essence of the discussed model  are the deterministic 
differential equations and not the calculation of probabilities as in 
the case of exponential model,  Poisson model, multinomial model or Markov 
chain models.  The model is not of the kind of gravity 
models as the distance among the origins and destinations of migration is
not presented explicitly in the model. The model possesses
some features of  replicator models. In more detail the migrants in each 
country of the channel (except for the final destination country) may follow
one of the strategies: (a) to remain in the corresponding country without 
change  of their status: (b) to remain in the country and to change their 
status from migrants to non-migrants; (c) to move to the next country of 
the channel. Replication is connected to the migrants that follow strategy (a). 
The change  of strategy of the members of the other two classes of migrants 
may be treated as mutation. The changes of the number $x_i$ of migrants 
that have strategy to stay as migrants in the i-th country of the channel 
is given by an ordinary differential equation as in the case of replicator 
model. But in our model we do not consider the payoffs connected to the 
different strategies  explicitly. What is taken into account is that the countries that are closer to the  final destination country are more preferred  
by the migrants. This determines the direction of the movement of migrants (from entry country of  the channel to the final destination country of the channel).
\par
From the point of view of the  distribution of the migrants in the
countries of the channel our model is a case of an urn model. Indeed the 
countries of the channel may be considered as urns  and the migrants
may be considered as  balls of the same color. Urns are numbered and at each 
time step there is addition of a number 
of balls in the urn with number $0$. This addition  is proportional to the 
number of the balls in all urns. Then some balls are removed from the
$0$-th urn  (corresponding to migrants that change their status) and  
some balls are moved to the urn with number $1$. The same actions of removing 
and transferring balls are performed to all urns (urn by urn following the
increasing urn numbers). Continuous approximation is assumed next. The goal is 
to obtain  the asymptotic stationary distribution of the balls in the urns. 
This asymptotic stationary distribution is the truncated Waring distribution.  
\par
The difference between the Waring distribution 
and the truncated Waring distribution from the point of view of our model is that there is a concentration of substance (migrants) in the last cell (final
destination country) of the sequence of cells (countries). If the final destination country is attractive then this concentration may be very significant: almost all migrants will  go (and may remain) there.
\par
The parameters that govern the distribution of migrants in the countries that
form the channels are $\sigma$, $\alpha$, $\beta$ and $\gamma$. $\sigma$ is the "gate" parameter as it regulates the number of migrants that enter the channel.  $\sigma$ is responsible for the (exponential) growth of the number of migrants in the channel. If $\sigma$ is large then the number of migrants may increase
very fast and this can lead to problems in the corresponding countries. Because
of this the states may try to keep $\sigma$ at satisfactory small value (and to
decrease the illegal migration) by official boundary entrances and even by fences. We note the $\sigma$ participates in each term of the truncated Waring distribution. This means that the situation at the entrance of the migration
channel influences significantly the distribution of migrants in the countries
of the channel. If the gates are open then a flood of migrants may be observed.
\par
The parameter $\gamma$ regulates the "absorption" of the channel as it regulates
the change of the status of some migrants. They can settle in the corresponding country (may obtain citizenship); may die on the route through the channel, etc.
The large value of $\gamma$ may compensate the value of $\sigma$ and even may
lead to decrease of the number of migrants in the channel. The large value of
$\gamma$ may lead to integration problems connected to migrants if the integration capacity of the corresponding country is limited. 
We shall discuss this question elsewhere.
\par
The parameter $\alpha$ regulates the motion of the migrants from one country to the next country of the channel. Small value of $\alpha$ means that the way of the migrants through the channel is more difficult and because of this the migrants tend to concentrate in the entry country (and eventually in the second
country of the channel). The countries that are at the second half of the migration channel and especially the final destination country may try to decrease $\alpha$ by agreements that commit the entry country to keep the migrants on its territory. Any increase of $\alpha$ leads to increase of the
proportion of migrants that reach the second half of the migration channel and especially the final destination country.
\par
The parameter $\beta$ regulates the attractiveness of the countries along the migration channel. Large values of $\beta$ mean that the final destination country is very attractive for some reason. This increases the attractiveness of
the countries from the second half of the channel (migrants want more to reach these countries as in such a way the distance to the final destination country
decreases). If for some reason $\beta$ is kept at high value a flood of migrants may reach the final destination country which may lead to large logistic and other problems there. 
\par
The discussed model of migration channels helps us to understand the 
functioning of such  channels and especially the effect of
concentration of migrants in a popular final destination country.
In addition the discussed model may be used for evaluation of
different scenarios of distribution of migrants in the countries that
belong to a  migration channel. Let
us discuss one possible scenario about the West-Balkan migration route from
Greece to Germany. The countries from this route are Greece, Macedonia, Serbia,
Croatia, Slovenia, Austria and Germany. The question we want to answer is about
the number of migrants without change of the status that are expected to be
at the end of the period in each of the countries of the channel. 
The UN agency of refugees estimated that from 1.10.2015 till 14.11.2015 
(3/2 months) 200 000 migrants are distributed in the countries of this route. 
Let us assume that in 2016 (from March till November, i.e.,
for 9 months) the intensity of motion through this migration channel remains the
same. This means  that one may expect that  1 200 000 migrants will be 
distributed among the countries of the channel at the end of the period. 
Let us assume that  $\sigma = 0.005$ (which means that at the end of period 
around 6000  migrants enter the channel through entry country per unit time,
i.e., per day). Let us assume that $\alpha =0.025$ and $\beta=0.006$ (
which means that 2.5\% of the migrants move from country to country because of
reasons not connected to  popularity of the final destination country and the
popularity of the final destination country is such that about 5\% of the
migrants from the country next to the final destination country move to the
final destination country per day).
Then according to the discussed model the contribution of the West-Balkan
route to  the numbers of  migrants without status in the corresponding
countries will be approximately as follows:
Greece: around $200 \ 000$; Macedonia: around $139 \ 000$; Serbia: around $102 \
000$; Croatia: around $81\ 000$; Slovenia: around $ 63 \ 000$; Austria: around
$51 \ 000$ and Germany: around $564 \ 000$. Note that for some countries like Germany other migration channels may contribute too  and the total number
of migrants without status in these countries may be larger.
The quantitative information obtained by the discussed model may be helpful in addition to information from other sources and models, e.g., it  may be
used for estimation of different kinds of expenses connected with the  migrants.
In addition the model allows quick estimation of the distribution of migrants
for the case of other scenarios (other values of the parameters of the channel).
Thus the model may contribute to the process of management and control of
migration flows. 
\begin{appendix}
\section{Waring distribution: properties and derivation}
The Waring distribution (named after Edward Waring - a Lucasian professor of
Mathematics in Cambridge in the 18th century) is a probability distribution on non-negative integers (Irwin (1963, 1968), Diodato (1994))
\begin{equation}\label{ap1}
p_i = \rho \frac{\alpha_{(i)}}{(\rho + \alpha)_{(i+1)}}; \
\alpha_{(i)} = \alpha (\alpha+1) \dots (\alpha+i-1)
\end{equation}
Waring distribution may be written also as follows
\begin{eqnarray}\label{ap2}
p_0 &=& \rho \frac{\alpha_{(0)}}{(\rho + \alpha)_{(1)}} = \frac{\rho}{\alpha + \rho}
\nonumber \\
p_i &=& \frac{\alpha+(i-1)}{\alpha+ \rho + i}p_{i-1}.
\end{eqnarray}
The mean $\mu$ (the expected value) of the Waring distribution is
\begin{equation}\label{ap3}
\mu = \frac{\alpha}{\rho -1} \ \textrm{if} \ \rho >1
\end{equation}
The variance of the Waring distribution is
\begin{equation}\label{ap4}
V = \frac{\alpha \rho (\alpha + \rho -1)}{(\rho-1)^2(\rho - 2)} \
\textrm{if} \ \rho >2
\end{equation}
$\rho$ is called the tail parameter as it controls the tail of the Waring
distribution. This can be see from the relationship for Waring distribution
when $i \to \infty$. Then 
\begin{equation}\label{ap5}
p_i \approx \frac{1}{i^{(1+\rho)}}.
\end{equation}
which is the frequency form of the Zipf distribution (Chen (1980)).
\par
The Waring distribution contains as particular cases other interesting distributions. One example is the famous Yule distribution (called also
Yule-Simon distribution). In this case $\alpha \to 0$ and
the Waring distribution is reduced to the Yule-Simon distribution (Simon (1955))
\begin{equation}\label{ap6}
p(\zeta = i \mid \zeta > 0) = \rho B(\rho+1,i)
\end{equation}
where $B$ is the beta-function. Another example is the geometric distribution.
Let $\alpha = \sigma/\beta$ and $\rho = a/\beta$. Then we obtain the
following variant of the Waring distribution:
$p_0=a/(\sigma+a)$ and $p_i=[(\sigma+\beta(i-1))/(a+\sigma+\beta i)]p_{i-1}$. let us now set $\beta=0$ in this variant of the distribution. We obtain $p_i =[\sigma/(\sigma + a)]p_{i-1}$  and the corresponding distribution
\begin{equation}\label{ap7}
p(\zeta = i) = q(1-q)^i; \ q = \frac{a}{\sigma+a}
\end{equation}
is called geometric distribution (Frank (1962), Coleman (1964)).
\par
The Waring distribution is a distribution with a very long tail. Because
of this property the Waring distribution is very suitable to describe
characteristics of many systems from the areas connected to research on  biology and society.
The Waring distributions is obtained when the model discussed in the
main text is applied for the case of an infinite sequence of cells $N \to \infty$. In this case we have infinite  array of cells (boxes) indexed in 
succession by non-negative integers, i.e., the first cell has index $0$. 
We assume again  that there exists an amount $x$ of some substance that is
distributed among the cells. If $x_i$ is the amount of the substance in the
$i$-th cell then $x = \sum \limits_{i=0}^\infty x_ i$. We introduce
$y_i = p(\zeta = i), \ i=0,1, \dots$ and assume that the expected value of the random variable $\zeta$ is finite $E(\zeta) = \sum \limits_{i=0}^\infty i y_i$.
\par
The content $x_i$ of any cell can change due to the same 3 processes as in the main text: (i) Some amount $s$ of the substance $x$  may enter the system of
cells from the external environment through the $0$-th cell; (ii) Amount $f_i$ can be transferred from the $i$-th cell into the $i+1$-th cell; (iii) Amount $g_i$  may leak out the $i$-th cell into the external environment.
These processes can be modeled mathematically by the system of ordinary
differential equations:
\begin{eqnarray} \label{war4}
\frac{dx_0}{dt} &=& s-f_0 - g_0; \nonumber \\
\frac{dx_i}{dt} &=& f_{i-1} -f_i - g_i, \ i=1,2,\dots
\end{eqnarray}
The following relationships for the amount of the moving substances may be assumed (Schubert and Gl{\"a}nzel (1984))
($\alpha, \beta, \gamma, \sigma$ are constants):
$s = \sigma x$; $\sigma > 0$ (self-reproducing property); 
$f_i = (\alpha + \beta i) x_i$; $\alpha >0$, $\beta \ge 0$ (cumulative 
advantage of higher cells); $g_i = \gamma x_i$; $\gamma \ge 0$  (uniform leakage over the cells). The substitution of these conditions in Eqs.(\ref{war4}) leads to the relationships
\begin{eqnarray}\label{war6}
\frac{dx_0}{dt} &=& \sigma x - \alpha x_0 - \gamma x_0; \nonumber \\
\frac{dx_i}{dt} &=& [\alpha+ \beta(i-1)]x_{i-1} - (\alpha + \beta i +\gamma)x_i
\end{eqnarray}
\par
Let us sum the equations from (\ref{war6}). The result of the summation is
$\frac{dx}{dt} = (\sigma - \gamma) x$ and the solution for $x$ is
$x = x(0) \exp[(\sigma - \gamma)t]$
where $x(0)$ is the amount of $x$ at $t=0$. 
\par
The distribution of $y_i$ will lead us to the Waring
distribution. From Eqs.(\ref{war6}) and with the help of the relationship
for $\frac{dx}{dt}$
and the relationship $\frac{dy_i}{dt} = \frac{1}{x^2}\left[ x \frac{dx_i}{dt}  -
x_i \frac{dx}{dt}\right]$ one obtains
\begin{eqnarray}\label{war8}
\frac{dy_0}{dt}&=& \sigma -(\alpha + \sigma) y_0; \nonumber \\
\frac{dy_i}{dt}&=&[\alpha+\beta(i-1)]y_{i-1} - (\alpha + \beta i + \sigma) y_i
\end{eqnarray}
The solution of Eq.(\ref{war8}) is
\begin{equation}\label{war9}
y_i = y_i^* + \sum \limits_{j=0}^i b_{ij} \exp[-(\alpha + \beta j + \sigma)t]
\end{equation}
where $y_i^*$ is the stationary solution of Eqs.(\ref{war9}) given
by the relationships
\begin{eqnarray}\label{war10}
y_0^* &=& \frac{\sigma}{\sigma + \alpha} \nonumber \\
y_i^* &=& \frac{\alpha + \beta(i-1)}{\alpha + \beta i + \sigma} y_{i-1}^*, \ i=1,2,\dots
\end{eqnarray}
Above $b_{ij}$ are determined by the initial conditions. In the exponential
function there are no negative coefficients and because of this when $t \to
\infty$  the sum in Eq.(\ref{war9}) vanishes and the systems comes to the
stationary distribution from Eqs.(\ref{war10}). This distribution is called Waring distribution. The explicit form of the Waring distribution is:
\begin{equation}\label{war11}
P(\zeta = i) = \frac{a}{a+k} \frac{(k-1)^{[i]}}{(a+k)^{[i]}}; \ k^{[i]} = \frac{(k+i)!}{k!}; \ i=0,1,\dots
\end{equation}
with parameters $k = \alpha/\beta$ and $a=\sigma/\beta$.
\end{appendix}
\vskip1cm
\begin{flushleft}
\textbf{\large References}
\end{flushleft}
\begin{description}
\item[]
Armitage, R., 1986. Population projections for English local authority areas, Population Trends \textbf{43} (Spring), 31-40.
\item[]
Bierley M.J.,Forster J.J., McDonald J.W., Smith P.W.F., 2008,
Bayessian estimation of migration flows, p.p. 149 - 174 in Raymer J.,
Willekens F., (Eds.). International migration in Europe: Data, models and 
estimates. Wiley, New York.
\item[]
Blom, G., Holst L., Sandell D., 1994, Problems and snapshots from the world 
of probability. Springer, Berlin.
\item[]
Blossfeld, H.-P., Rohwer G., 2002, Techniques of event history modeling: new approaches to casual analysis. Lawrence Erlbaum, New Jersey.
\item[]
Blossfeld, H.-P., Golsch K., Rohwer G. (Eds.), 2007, Event history analysis 
with Stata . Lawrence Erlbaum, New Jersey.
\item[]
Bowley, A. L., 1924. Births and population change in Great Britain, The Economic Journal, \textbf{34},  188 - 192.
\item[]
Bracken, I., Bates J.J., 1983, Analysis of gross migration profiles in England and Wales: some developments in classification,  Environment and Planning A,
\textbf{15}  343-355.
\item[]
Cadwallader, M., 1989,  A conceptual framework for analysing migration behaviour in the developed world, Progress in Human Geography \textbf{13}, 494 - 511.
\item[]
Cadwallader, M., 1992, Migration and residential mobility. Macro and micro approaches, The University of Wisconsin Press, Madison, Wisconsin.
\item[]
Champion, A. G., Bramley, G., Fotheringham, A. S., Macgill, J.,  Rees, P. H., 2002, A migration modelling system to support government decision-making,  p.p. 257-278 in J. Stillwell, S. Geertman, (Eds.), Planning support systems in practice, Springer, Berlin.
\item[]
Chen, W. -C., 1980, On the weak form of the Zipf's law, Journal of Applied Probability \textbf{17}, 611 - 622
\item[]
Coleman, J. S., 1964, Introduction to mathematical sociology, Collier-Macmillan,
London.
\item[]
Collins L.,1972. Industrial migration in Ontario: forecasting
aspects of industrial activity through Markov chains. Statistics Canada, 
Ottawa.
\item[]
Collins L., 1975, An introduction to Markov chain analysis. Headey Brothers
Ltd., London.
\item[]
Congdon, P., 1991, An application of general linear modelling to migration in London and the South East,  p.p. 113 - 136 in  J. C. H. Stillwell, P. Congdon, (Eds.), Migration models: Macro and micro perspectives, Belhaven Press, London.
\item[]
Cressman R., Tao Y., 2014, The replicator equation and other game dynamics.
PNAS \textbf{111}, Suppl.3, 10810-10817.
\item[]
Diodato, V., 1994, Dictionary of bibliometrics, Haworth Press, Binghampton, New York.
\item[]
Flajolet P., Dumas, P., Puyhaubert V., 2006, Some exactly solvable models of
urn process theory. Discrete Mathematics and Theoretical Computer Science 
(DMTCS) proceedings, Nancy, France, 59 - 118.  
\item[]
Flowerdew, R., 1991, Poisson regression modelling of migration, p.p. 92 - 112 in J. C. H. Stillwell, P. Congdon, (Eds.), Migration models: Macro and micro approaches, Belhaven Press, London.
\item[]
Fotheringham, A. S., Nakaya, T., Yano, K., Openshaw, S., Ishikawa, Y., 2001, 
Hierarchical destination choice and spatial interaction modelling: a 
simulation experiment, Environment and Planning A \textbf{33}, 901 - 920.
\item
Fotheringham, A. S., Brunsdon, C., Charlton, M., 2002, Geographically weighted
regression: The analysis of spatially varying relationships, Wiley, Chichester.
\item[]
Frank, R., 1962, Brand choice as a probability process, Journal of Business
\textbf{35}, 43 - 56.
\item[]
Greenwood, M.J., 2005. Modeling migration, p.p. 725 - 734 in Kemp-Leonard, K. (Ed.)  Encylopedia of social measurement, vol. 2, Elsevier, Amsterdam.
\item[]
Hachen D.S., 1988. The competing risk model. Sociological Methods and Research \textbf{17}, 21 - 54. 
\item[]
Irwin, J. O., 1963, The place of mathematics in medical and biological sciences, Journal of the Royal Statistical Society \textbf{126} (1963) 1 - 44.
\item[]
Irwin, J. O., 1968, The generalized Waring distribution applied to accident theory, Journal of the Royal Statistical Society \textbf{131} (1968) 205 - 225. 
\item[]
Karev, G.P., Kareva, I.G., 2014, Replicator equations and models of biological
populations and communities. Math. Model. Nat. Phenom. \textbf{9}, 68 - 95.
\item[]
Ledent J., 1980, Multistate life table: movement versus transition perspectives.
Environment and Planning A \textbf{12}, 533 - 562.
\item[]
McLachlan G.J.,  Krishnan T., 1997. The EM algorithm and extensions. Wiley, 
New York.
\item[]
Maier, G., Weiss, P., 1991, The discrete choice approach to migration modelling,
p.p. 17 - 31 in J. C. H. Stillwell, P.  Congdon, (Eds.), Migration models: Macro and micro approaches, Belhaven Press, London.
\item[]
Nakaya, T., 2001,  Local spatial interaction modelling based on the geographically weighted regression approach, GeoJournal \textbf{53} (2001) 347 - 358.
\item[]
Ravenstein, E. G., 1885, The laws of migration, Journal of the Royal Statistical Society, \textbf{48}(II),  167 - 227.
\item[]
Raymer J., 2007, The estimation of international migration flows: a 
general technique focused on the origin-destination association structure.
Environment and Planning A \textbf{39}, 985 - 995.
\item[]
Rees, P. H., Wilson, A. G., 1977, Spatial population analysis, Edward Arnold, London.
\item[]
Rogers, A., 1990, Requiem for the net migrant. Geographical Analysis \textbf{22}, 283 - 300.
\item[]
Rogers, A., 1995, Multiregional demography, Wiley, London.
\item[]
Schubert, A., Gl{\"a}nzel, W., 1984, A dynamic look at a class of skew distributions. A model with scientometric application, Scientometrics \textbf{6}, 149 -- 167.
\item[]
Simon, H. A., 1955, On a class of skew distribution functions, Biometrica \textbf{42}, 425 - 440. 
\item[]
Singer B., Spilerman S. 1979. Mathematical representations of 
development. Theories. p.p. 155 - 177 in Nesselroade J.R.,  Baltes P.B. (Eds.)
Longitudinal research in the study of behavior and 
development.Academic Press, New York.
\item[]
Stillwell, J. C. H., 1978, Interzonal migration: some historical tests of spatial interaction models, Environment and Planning A \textbf{10}, 1187 - 1200.
\item[]
Stillwell, J. C. H., Congdon P.,(Eds.), 1991, Migration models: Macro and micro
approaches, Belhaven Press, London.
\item[]
Vitanov, N. K., Dimitrova, Z. I., Ausloos M., 2010, Verhulst-Lotka-Volterra model of ideological struggle, Physica A \textbf{389}, 4970 - 4980.
\item[]
Vitanov, N. K., Ausloos, M., Rotundo, G., 2012, Discrete model of ideological struggle accounting for migration, Advances in Complex Systems \textbf{15}, Supplement 1, Article number 1250049.
\item[]
Vitanov N. K., Jordanov, I. P., Dimitrova, Z. I., 2009, On nonlinear dynamics of interacting populations: Coupled kink waves in a system of two populations,
Communications in Nonlinear Science and Numerical Simulation \textbf{14}, 2379 - 2388.
\item[]
Vitanov, N. K., Jordanov, I. P., Dimitrova, Z. I., 2009, On nonlinear population waves, Applied Mathematics and Computation \textbf{215}, 2950 - 2964.
\item[]
Vitanov, N.K., Dimitrova, Z. I., 2010, Application of the method of simplest equation for obtaining exact traveling-wave solutions for two classes of model PDEs from ecology and population dynamics. Communications in Nonlinear Science
and Numerical Simulation, \textbf{15}, 2836 - 2845.
\item[]
Vitanov, N. K., Dimitrova, Z. I., Vitanov, K. N., 2011, On the class of nonlinear PDEs that can be treated by the modified method of simplest equation. Application to generalized Degasperis - Processi equation and b-equation.
Communications in Nonlinear Science and Numerical Simulations \textbf{16}, 3033 - 3044.
\item[]
Vitanov, N. K., Dimitrova, Z. I., Kantz, H., 2013, Application of the method of simplest equation for obtaining exact traveling-wave solutions for the extended Korteweg - de Vries equation and generalized Camassa - Holm equation. Applied
Mathematics and Computation \textbf{219}, 7480 - 7492.
\item[]
Vitanov, N. K,  Dimitrova, Z. I., Vitanov, K. N., 2013, Traveling waves and statistical distributions connected to systems of interacting populations. Computers \& Mathematics with Applications \textbf{66}, 1666 - 1684.
\item[]
Vitanov, N. K., Vitanov, K. N., 2014, Population dynamics in presence of state dependent fluctuations. Computers \& Mathematics with Applications \textbf{68}, 962 - 971.
\item[]
Vitanov, N. K., Dimitrova, Z. I., 2014, Solitary wave solutions for nonlinear partial differential equations that contain monomials of odd and even grades with respect to participating derivatives. Applied Mathematics and Computation
\textbf{247}, 213 - 217.
\item[]
Vitanov, N. K., Dimitrova, Z. I., Vitanov, K. N., 2015, Modified method of simplest equation for obtaining exact analytical solutions of nonlinear partial differential equations: further development of the methodology with applications. Applied Mathematics and Computation \textbf{269}, 363 - 378
\item[]
Whelpton, P. K., 1936, An empirical method of calculating future population, Journal of the American Statistical Association \textbf{31}, 457 - 473.
\item[]
Willekens F.J., 1999, Probability models of migration: Complete and incomplete data. SA Journal of Demography \textbf{7}, 31 - 43.
\item[]
Willekens, F., 2008, Models of migration observations and judgement. p.p. 117 - 
147 in J. Raymer, F. Willekens (Eds.) International migration in Europe: Data, 
models and estimates. Wiley, New York.
\item[]
Wilson, A. G., 1970, Entropy in urban and regional modelling, Pion, London.
\item[]
Zipf, G. K., 1946, The P1P2/D hypothesis: on intercity movement of persons, American Sociological Review \textbf{11}, 677 - 686.
\end{description}

\end{document}